\documentstyle[aps,manuscript,psfig]{revtex}
 
\begin{document}
 
\draft

\title{Irreversible phase transitions induced by an oscillatory input.}

\author{A. C. L\'{o}pez\footnote{ e-mail: ac-lopez@yahoo.com },
 G. P. Saracco\footnote{ e-mail: gsaracco@inifta.unlp.edu.ar }
 and E. V. Albano\footnote{ e-mail: ealbano@inifta.unlp.edu.ar } }

\address{Instituto de Investigaciones Fisicoqu\'{\i}micas Te\'{o}ricas y Aplicadas
(INIFTA), Facultad de Ciencias Exactas, Universidad Nacional de La Plata (CONICET,
CIC, UNLP), Sucursal 4, Casilla de Correo 16, (1900) La Plata, Argentina.}

\maketitle
\begin{abstract}
A novel kind of irreversible phase transitions (IPT's) driven by an oscillatory
input parameter is studied by means of computer simulations. Second order IPT's
showing scale invariance in relevant dynamic critical properties are found to
belong to the universality class of directed percolation. In contrast, the absence
of universality is observed for first order IPT's. 
\end{abstract}

Far from equilibrium system often exhibits irreversible phase transitions (IPT's)
between an active (or reactive) regime and an inactive (or absorbing) state.
Such transitions are irreversible because a system trapped in an absorbing state
can never escape from it. Among others, models exhibiting IPT's are directed
percolation\cite{ref1}, contact processes\cite{ref2}, branching annihilating
walkers\cite{ref3}, forest-fire models \cite{ref4,g22,g24}, models for dynamic
evolution of living individuals \cite{ref5} and several models of catalyzed
reactions such as the ZGB model \cite{ref6} and variations \cite{ref10,ref11}.

In spite of the considerable progress achieved in the understanding of irreversible
critical behavior (for reviews see e. g. \cite{ref10,ref11,xref2}), the study
of the dynamic response of irreversible systems to external perturbations is
still in its infancy. Very recently, we have studied the dynamic response of
the ZGB model close to a second order IPT, showing that, after driving the system
within the absorbing state, the subsequent relaxation can be well described
by an stretched exponential decay\cite{ref14}. The lack of additional studies
in this field is in contrast to their equilibrium counterpart. In fact, the
study of dynamic response systems in thermodynamic equilibrium, close to reversible
phase transitions, to an external perturbation, is a subject of current interest
\cite{ref15}.

The aim of this work is to study the dynamic response of irreversible systems
to a periodic external perturbation. The study has been performed close to both
first and second order IPT's. For this purpose we have selected a forest fire model
with immune trees (FFMIT)\cite{g22} and the Ziff-Gulari-Barshad model (ZGB
model ) \cite{ref6}.

The ZGB model is a lattice gas reaction system aimed to mimic the catalytic
oxidation of carbon monoxide, \( CO+(1/2)O_{2}\rightarrow CO_{2} \), according
to the Langmuir-Hinshelwood mechanism. So, reactants \( CO \) and \( O_{2} \)
are adsorbed on the surface of the catalyst with probabilities \( P_{CO} \)
and \( P_{O_{2}} \). Since these probabilities are normalized ( \( P_{CO}+P_{O_{2}}=1 \)),
the ZGB model has a single parameter, i.e. \( P_{CO} \). For \( P_{CO}\rightarrow 1 \)
( \( P_{CO}\rightarrow 0 \) ), the surface of the catalyst becomes inactive
due to complete saturation with \( CO \) ( \( O \) ) species, respectively.
However, between this two inactive states there is a reactive regime as it is
shown in Figure \ref{ZGBdiagram}. Close to \( P_{2\, CO}\simeq0.5256 \) the
ZGB exhibits an abrupt change in the rate of reaction and reactant's coverages,
as it is shown in Figure \ref{ZGBdiagram}, indicating a first order IPT. Further
details on the ZGB model can be found elsewhere \cite{ref10}.

In order to study the dynamic response of the system, first a stationary configuration
of the ZGB model well inside the reactive regime (actually in the center of
the reaction window \( P^{w}_{CO}=0.455 \) ) is obtained. Subsequently, an
square oscillatory perturbation of the form

\begin{equation}
\label{eqpco}
P_{CO}\left( t\right) =\left\{ \begin{array}{ll}
P_{CO}^{w}+A_{P_{CO}} & ,\textrm{ if }0\leq t<\tau /2\\
P_{CO}^{w} & ,\textrm{ if }\tau /2\leq t<\tau \\
P_{CO}(t)=P_{CO}(t+\tau ) & ,\, \, \forall \textrm{ }t\geq 0
\end{array}\right. 
\end{equation}
 is applied to the system, where \( A_{P_{CO}} \) is the amplitude and \( \tau  \)
is the period of the perturbation ( see also Figure \ref{ZGBdiagram}).

There is a great variety of forest-fire models (FFM's), e. g.\cite{ref4,g22,g24},
for an extensive review see e. g. \cite{g29}. FFM's are stochastic cellular
automata which are defined on \( d \)-dimensional hypercube lattices with \( L^{\, d} \)
sites. Each site can be either occupied by a tree, a burning tree, or empty.
A great number of FFM's can be defined giving a set of rules which are used,
during each time step, to update the system in parallel. These rules are: (1)
burning tree \( \rightarrow  \) empty site; (2) tree \( \rightarrow  \) burning
tree with probability \( (1-g) \) if at least one nearest neighbour is burning;
and (3) empty site \( \rightarrow  \) tree with probability \( p \). The probability
\( p \) is to be taken as growing probability and \( g \) is the immunity
of each tree to catch fire.

In the present work we shall focus our attention to the forest-fire model with
immune trees (FFMIT)\cite{ref4,g22,g24}. Qualitatively speaking, if the growing
probability is non-zero (\( p\rightarrow 1 \)) and the immunity is low (\( g\rightarrow 0 \))
one expects coexistence of fire, trees and empty sites. However, keeping \( p \)
constant and increasing \( g \) the fire will eventually cease and the system
will become irreversibly trapped in an absorbing state with the lattice completely
filled by trees. So, the FFMIT exhibits second order IPT's between an active
state with fire propagation and an absorbing state where the fire becomes irreversibly
extinguished \cite{ref4,g24}.

In order to study the dynamic response of the FFMIT upon temporal variations
of the parameters, one can vary either \( p \), \( g \), or both of them.
However, we have worked taking \( p=constant \) while \( g \) is varied. For
this purpose, the procedure is as follows: first a stationary active state of
the standard FFMIT is obtained for fixed values of the parameters. In this work
we take \( p=p_{0}=0.5 \) and \( g=g_{0}=0.46 \). Subsequently \( p \) is
kept fixed and \( g \) is varied harmonically according to

\begin{equation}
\label{ec1FFMIT}
g=\left( g_{0}+\frac{A_{g}}{2}\right) +\frac{A_{g}}{2}\, sin(\frac{2\pi }{T}t)\, ,
\end{equation}
 where \( A_{g} \) and \( T \) are the amplitude and the period of the oscillation,
respectively. Notice that the critical point is given by \( p_{0}=p_{c}=0.5 \)
and \( g_{c}=0.5614\pm 0.0005 \) \cite{ref4,g24}.

Due to the variation of the parameters, either \( g(t) \) or \( P_{CO}(t) \),
it is expected that for long periods and/or large amplitudes, the systems may
eventually become trapped into an absorbing state. So, the oscillatory variation
of the parameters may cause IPT's from the active (oscillatory) regimes to the
absorbing states. These transitions may occur at critical values of the amplitude
(\( A^{c}_{g} \) and \( A^{c}_{P_{CO}} \) ) and the period (\( T_{c} \) and
\( \tau _{c} \) ). In order to characterize and study such IPT's, we have performed
epidemic simulations (ES) \cite{ref1,ref2,ref3,ref4}. The idea behind ES
is to start from a configuration very close to the absorbing state and subsequently,
to follow the temporal evolution of the system under consideration. Therefore,
we took a sample filled with trees (or \( CO \) molecules) except for a small
patch of \( 2\times 2 \) sites having burning trees (or empty sites) placed
at the center of the lattice for the FFMIT and the ZGB model, respectively.
Depending on the values of the parameters and due to the stochastic nature of
the processes, such a small perturbation would either propagate, or eventually
become extinguished. During the epidemic propagation the following quantities
are measured: (i) the average number of burning trees (empty sites) \( N(t) \),
respectively; (ii) the survival probability \( P(t) \), i.e. the probability
that the fire is still ignited (there are empty sites) at time \( t \), respectively;
and (iii) the average mean-square distance \( R^{2}(t) \) over which the fire
(empty sites) has spread, respectively.

The usual ansatz for ES close to second order IPT's is to assume that \( N(t) \),
\( P(t) \) and \( R^{2}(t) \) would obey power-law dependencies with exponents
\( \eta  \),\( -\delta  \) and \( z \)\cite{ref1}, respectively. However,
in the present ES of the FFMIT the input parameter varies harmonically, so we
expect to obtain an oscillatory output modulated by a power-law, that is 
\begin{equation}
\label{eqNtEsA}
N(t)\sim t^{\eta }.cos(\frac{2\pi }{T}t+B)\, ,
\end{equation}
 and similarly for \( P(t) \) and \( R^{2}(t) \), where \( B \) is a constant
phase-shift.

Figure \ref{gfig3}(a) shows a \( log-log \) plot of \( N(t) \) vs. \( t \)
obtained performing ES, where the influence of oscillatory input can clearly
be observed. In order to perform a fit, we have first determined the values
of \( N(t) \) on peaks, valleys and centers, given by \( N^{+} \), \( N^{-} \)and
\( N^{0} \), respectively. The insert of figure \ref{gfig3}(a) shows that \( log-log \)
plots of \( N^{+} \), \( N^{-} \)and \( N^{0} \) versus \( t \) can be very
well fitted by straight lines with slopes \( \eta ^{+}=0.218 \) , \( \eta ^{-}=0.223 \)
, \( \eta ^{0}=0.224 \), respectively. As it is usual, plots drawn taking smaller
(larger) amplitudes show upward (downward) curvature suggesting that they are
off-criticality, respectively. Precisely, the straight line observed is the signature
of a power-law behavior which characterizes a second-order phase transition
exhibiting scale invariance. The insert in Figure \ref{gfig3}(b) shows that
it is also possible to determine the assumed constant phase shift in equation
(\ref{eqNtEsA}), comparing plots of \( N(t) \) and \( g(t) \) versus \( t \).
For example, in Figure \ref{gfig3}(b), we have obtained \( B=1.37\pm 0.34 \)
. Therefore we can fit the whole curve of \( N(t) \) vs. \( t \) using the
already determined values of \( \eta  \) and \( B \), as it is shown in Figure
\ref{gfig3}(b).

Figure \ref{gfig4} shows a \( log-log \) plot of \( P(t) \) versus \( t \)
which also exhibits oscillatory behavior. In this case, the drops of the survival
probability are due to the increment in the immunity which drives the system
into the absorbing state making the fire propagation harder, an effect which
may cause the eventual extinction of some epidemics. Defining the maximum, medium
and minimum values of \( P(t) \) in each cycle as \( P^{+} \), \( P^{-} \)and
\( P^{0} \), respectively, we can fit the exponent \( \delta  \), as it is
shown in the insert of figure \ref{gfig4}. Our results, at criticality, are
\( \delta ^{+}\cong 0.463 \) , \( \delta ^{-}\cong 0.400 \) , \( \delta ^{0}\cong 0.433 \)
respectively.

It is found that \( R^{2} \) is less sensitive to the oscillatory input than
\( N(t) \) and \( P(t) \). The plot of \( R^{2} \) vs. \( t \) can roughly
be fitted by a straight line which yields a slope \( z\cong 1.18 \) (not shown
here for the sake of space).

It should be noticed that the dynamic exponents which characterize the IPT's
driven by the oscillatory parameter are in good agreement with those of the
universality class of directed percolation (DP, in 2+1 dimensions), namely \( \eta =0.214 \),
\( \delta =0.460 \), and \( z=1.134 \) \cite{ref1}. Also, the hyperscaling
relation \( d\, z=4\, \delta +2\, \eta  \) \cite{ref1}, is rather well satisfied
by these exponents. So, we conclude that the novel type of transition discussed
so far, can be placed in the universality class of DP. This result extends the
validity of Janssen's conjecture \cite{g31}, which states that a continuous
transition into an absorbing state characterized by an scalar order parameter
may belong to the universality class of DP, to second order irreversible transitions
driven by oscillatory parameters.

Performing ES with different values of \( A_{g} \) and \( T \) we have evaluated
the phase diagram of the FFMIT under oscillatory driving, as it is shown in
figure \ref{FFMITosc}. The critical curve \( T_{c} \) vs. \( A_{g_{c}} \)
shows the location of second order IPT's between the active regime (trees+burning
trees+empty sites) and the absorbing state (only trees). All of these transitions
belong to the universality class of DP. The insert of Figure \ref{FFMITosc}
shows a \textit{log-log} plot of \( T_{c} \) vs. \( \Delta A=A_{gc}-(g_{c}-g_{0}) \).
Notice that \( \Delta A \) is the ``excess critical amplitude'', namely a
renormalized amplitude which accounts for the value of the oscillatory parameter
which exceeds the stationary critical threshold allowing the system to make
an excursion to the absorbing state. The data is then consistent with an hyperbolic-like
behavior of the form \( T_{c}\propto \Delta A^{\alpha } \) with exponent \( \alpha \cong 4.55 \).
The deviation from this behaviour, observed for short periods, is likely due
to the fact that the input signal can not longer be considered as harmonic.

Figure \ref{alfig4} shows \( log-log \) plots of \( N(t) \) versus \( t \)
obtained performing ES of the ZGB model for a fixed period (\( \tau =20\, MCS \)
) and different values of the amplitude \( A_{P_{CO}} \). In contrast to the FFMIT,
here \( N(t) \) decreases with time as it has been observed for ES at first
order IPT's \cite{evans36}. Figure \ref{alfig4} also allows to identify subcritical
(\( A_{P_{CO}}=0.123 \)), supercritical (\( A_{P_{CO}}=0.125 \)) and critical
(\( A^{c}_{P_{CO}}=0.12345 \)) behavior, respectively. In contrast to the ES
of the second order IPT's of the FFMIT which exhibit power-law behavior (Figure
\ref{gfig3}), the plot of figure \ref{alfig4}, at criticality, exhibit marked
curvature suggesting the cross over to a cut-off at certain long time. Based
on the observation that the short time behavior of \( N(t) \) vs. \( t \)
seems to obey a pseudo power-law with effective exponent \( \eta ^{eff}\cong 2 \),
we have proposed \cite{37} the following ansatz as a generalization of equation
(\ref{eqNtEsA}) 
\begin{equation}
\label{eqNtexpZGB}
N(t)\propto (t/\tau ^{*})^{-\eta ^{eff}}exp(-t/\tau ^{*})\, ,
\end{equation}
 where \( \tau ^{*} \) set a characteristic time scale and the oscillatory
dependence of \( N(t) \) is not considered for the sake of simplicity. Measuring
the value of the average number of holes at half period \( N^{*}(t) \) it is
possible to check the validity of the ansatz given by equation (\ref{eqNtexpZGB}).
In fact, the insert of Figure \ref{alfig4} shows a \textit{semilogarithmic}
plot of \( N(t)\, (t/\tau ^{*})^{\eta ^{eff}} \) vs. \( (t/\tau ^{*}) \) can
be well fitted with a single parameter, namely the characteristic time \( \tau ^{*}\simeq 471\pm 5 \).
This results indicates that the power-law behavior for short times (\( t<\tau ^{*} \))
crosses over to an exponential decay (\( t>\tau ^{*} \)). Our results not only
point out that the IPT's driven by the oscillatory parameter in the ZGB model
are of first order, but also allow us to rule out the occurrence of power law
(scale invariance) behavior. This finding conciliates the behavior of first
order IPT's with their equilibrium counterpart where it is well know that the
existence of short range correlations prevents the occurrence of scale invariance.

Performing ES of the ZGB model under oscillatory driving of the input parameter
for different values of the period we have evaluated the corresponding phase
diagram, namely a plot of \( \tau _{c} \) vs. \( A_{P_{CO}}^{c} \) , as it
is shown in Figure \ref{ZGBosc}. The critical curve shows the location of first
order IPT's between a reactive state with \( CO_{2} \) production and a poisoned
inactive state where the catalyst's surface is fully covered by \( CO \). The
insert of Figure \ref{ZGBosc} shows a \textit{log-log} plot of \( \tau _{c} \)
vs. \( \Delta A=A_{P_{CO}}^{c}-(P_{2\, CO}-P^{w}_{CO}) \). As in the previous
case of the FFMIT (Figure \ref{FFMITosc}) the data is consistent with an hyperbolic-like
behavior \( \tau _{c}\propto \Delta A^{\alpha } \) with exponent \( \alpha \cong 1.53 \).
It is also interesting to know that for short periods (\( \tau _{c}<1 \) ),
the value of the excess amplitude saturates just at \( \Delta A_{s}=A_{P_{CO}}^{c}/2=P_{2\, CO}-P^{w}_{CO}=0.071 \).
This result can straightforwardly be interpreted assuming that, such short
periods are indeed much shorter than the relaxation time and consequently the
system only feels the average value of the applied oscillatory pressure. Under
these circumstances, the critical amplitude is expected to be \( A_{P_{CO}}^{c}=P^{w}_{CO}+2.\Delta A_{s} \),
as it has already been observed in Figure \ref{ZGBosc}.

In summary, IPT's induced by an oscillatory external parameter are studied in
two different systems, namely a forest fire model with immunity and a model
for the catalytic oxidation of carbon monoxide. Second order IPT's are placed
in the universality class of directed percolation. However, first order IPT's
lacks of universal behavior. These findings are in qualitative agreement with
well established concepts developed in the study of reversible phase transitions.
A phase diagram for this new type of IPT's is a critical curve, \( period \)
versus \( amplitude \) of the oscillations which sets the boundary, between
an active regime and an absorbing state. These kinds of phase diagrams are evaluated
for both studied models and, in both cases, hyperbolic-like critical curves
are found but the corresponding exponents are different.

We expect that these findings will stimulate theoretical and experimental work
on the field of far from equilibrium systems in general and in the study of
irreversible critical behaviour in particular.

\acknowledgements 
This work was supported by CONICET, UNLP, ANPCyT, Fundaci\'{o}n Antorchas (Argentina),
and Volkswagen Foundation (Germany).

\begin{figure}

\caption{Plots of \protect\( R_{CO_{2}}\protect \), \protect\( \theta _{CO}\protect \)
and \protect\( \theta _{O}\protect \) for the ZGB model on the square lattice
of side \protect\( L=512\protect \). In the upper part the double arrow shows
the range where the \protect\( CO\protect \) pressure is oscillatory varied.
More details in the text.}

\label{ZGBdiagram}
\end{figure}
\begin{figure}

\caption{Results for the FFMIT obtained performing ES. (a) \protect\( Log-log\protect \)
plot of \protect\( N(t)\protect \) vs. \protect\( t\protect \) with \protect\( A_{g}=0.1695\protect \)
and \protect\( T=55\protect \). The insert shows \protect\( Log-log\protect \)
plots of \protect\( N^{+}\protect \), \protect\( N^{-}\protect \)and \protect\( N^{0}\protect \)
versus \protect\( t\protect \) (see text). (b) Fit of \protect\( N(t)\protect \)
vs. \protect\( t\protect \) (see text). The insert shows the constant phase
shift \protect\( B\protect \) between input and the output signal.}

\label{gfig3}
\end{figure}
\begin{figure}

\caption{\protect\( Log-log\protect \) plot of \protect\( P(t)\protect \) versus \protect\( t\protect \)
obtained performing ES for the FFMIT with \protect\( A_{g}=0.1695\protect \)
and \protect\( T=55\protect \). The insert shows the linear fits giving \protect\( \delta ^{+}\protect \),
\protect\( \delta ^{-}\protect \) and \protect\( \delta ^{0}\protect \)(for
more details see the text).}

\label{gfig4}
\end{figure}
\begin{figure}

\caption{Critical curve \protect\( T_{c}\protect \) vs. \protect\( A_{g_{c}}\protect \)
showing the location of second order IPT's driven by the oscillatory input between
the active regime (\protect\( AR\protect \), trees+burning trees+empty sites)
and the absorbing state (\protect\( AS\protect \), only trees) in the FFMIT.
The insert shows a \emph{log-log} of \emph{\protect\( T_{c}\protect \)} versus
\protect\( \Delta A=A_{g_{c}}-(g_{c}-g_{0})\protect \)\emph{.} The solid straight
line has slope \protect\( \alpha =4.55\protect \)\emph{.} More details in the
text. \emph{\label{FFMITosc}}}
\end{figure}
\begin{figure}

\caption{\protect\( Log-log\protect \) plots of \protect\( N(t)\protect \) versus
\protect\( t\protect \) obtained performing ES of the ZGB model for a fixed
period (\protect\( \tau =20\, MCS\protect \) ) and different values of \protect\( A_{P_{CO}}\protect \)
as indicated in the figure. The insert shows a semi logarithmic plot of \protect\( N(t)\left( t/\tau ^{*}\right) ^{\eta _{eff}}\protect \)
vs. \protect\( t/\tau ^{*}\protect \) . The straight line, which validates
the ansatz of the equation (\ref{eqNtexpZGB}) has been fitted to the values
of the output signal at half period \protect\( N^{*}(t)\protect \) , obtaining
\protect\( \tau ^{*}=471\pm 0.5\protect \) . More details in the text.}

\label{alfig4}
\end{figure}
\begin{figure}

\caption{Critical curve \protect\( \tau _{c}\protect \) vs. \protect\( A^{c}_{P_{CO}}\protect \)
, obtained for the ZGB model applying harmonic perturbations. The curve set
the boundary between the active regime (\emph{AR}) and the absorbing state (\emph{AS}).
The insert shows a \emph{log-log} of \protect\( \tau _{c}\protect \) versus
\protect\( \Delta A=A^{c}_{P_{CO}}-(P_{2\, CO}-P^{w}_{CO})\protect \). The
solid straight line has slope \protect\( \alpha =1.527\protect \). The vertical
dashed line shows the saturation of \protect\( \Delta A\protect \) for \protect\( \tau _{c}<1\protect \).
More details in the text. \label{ZGBosc}}
\end{figure}

{\par\centering \hbox{\psfig{figure=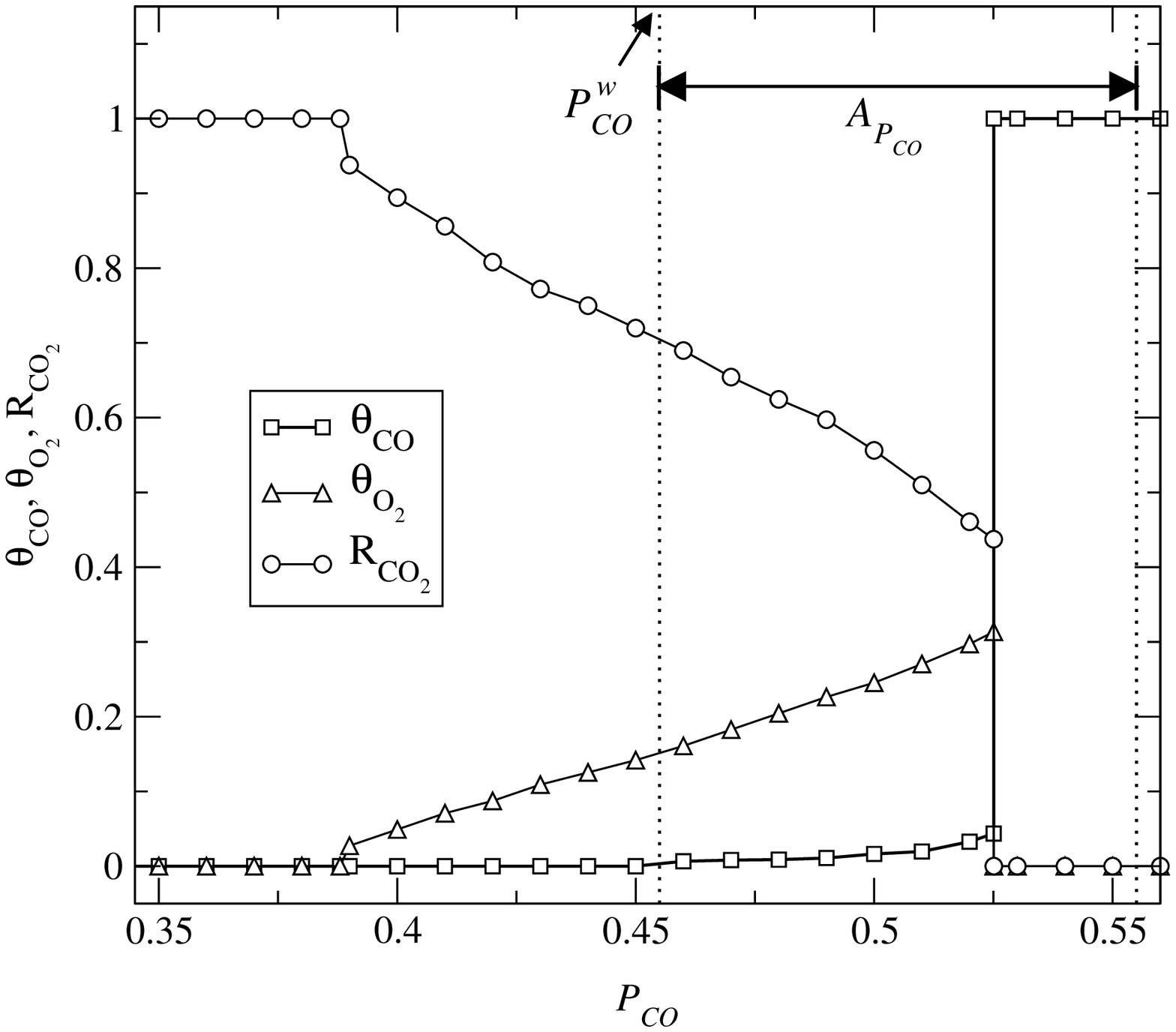,width=17cm}} Figure 1 \newpage
\hbox{\psfig{figure=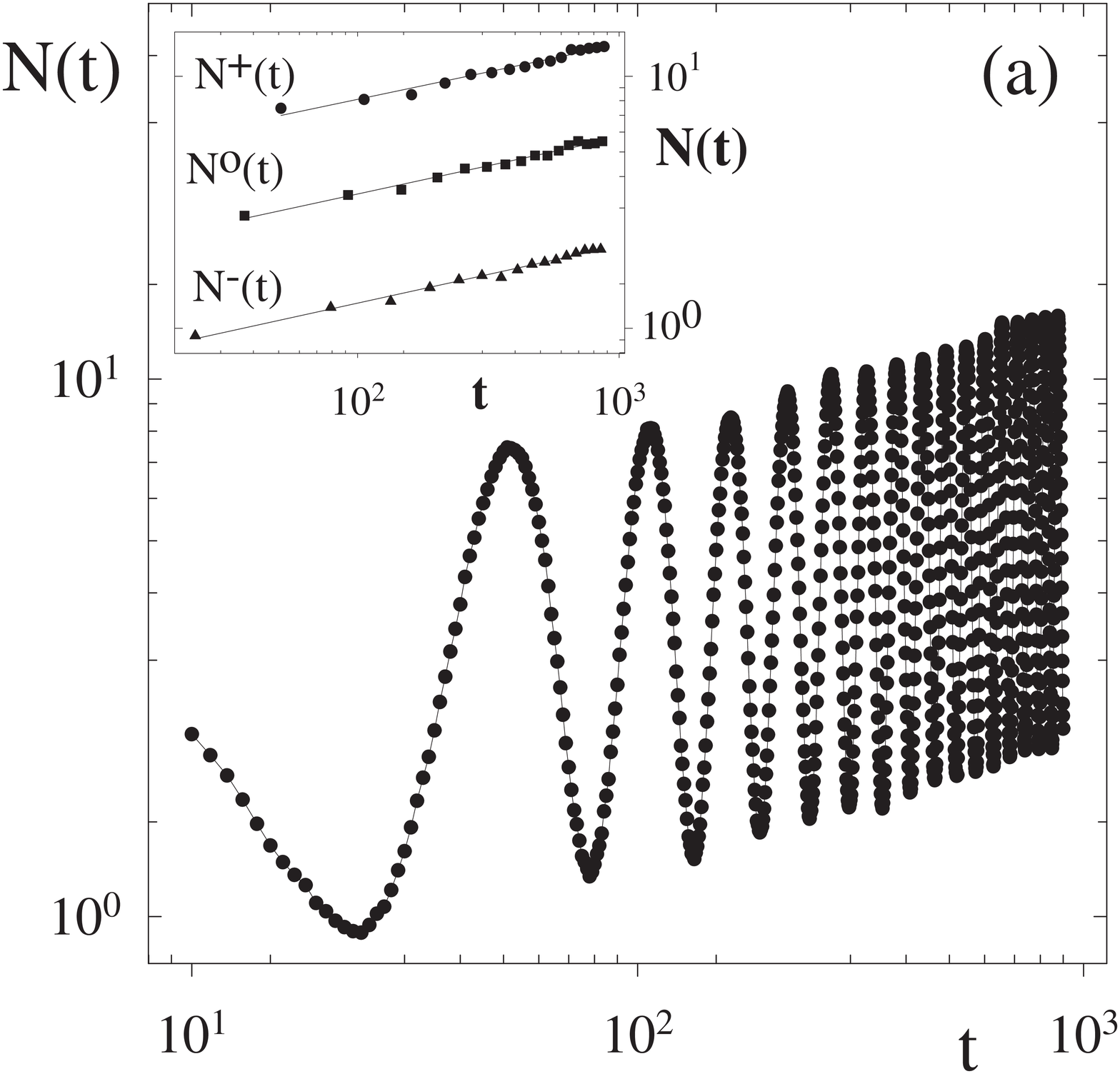,width=17cm}} Figure 2 (a) \newpage
\hbox{\psfig{figure=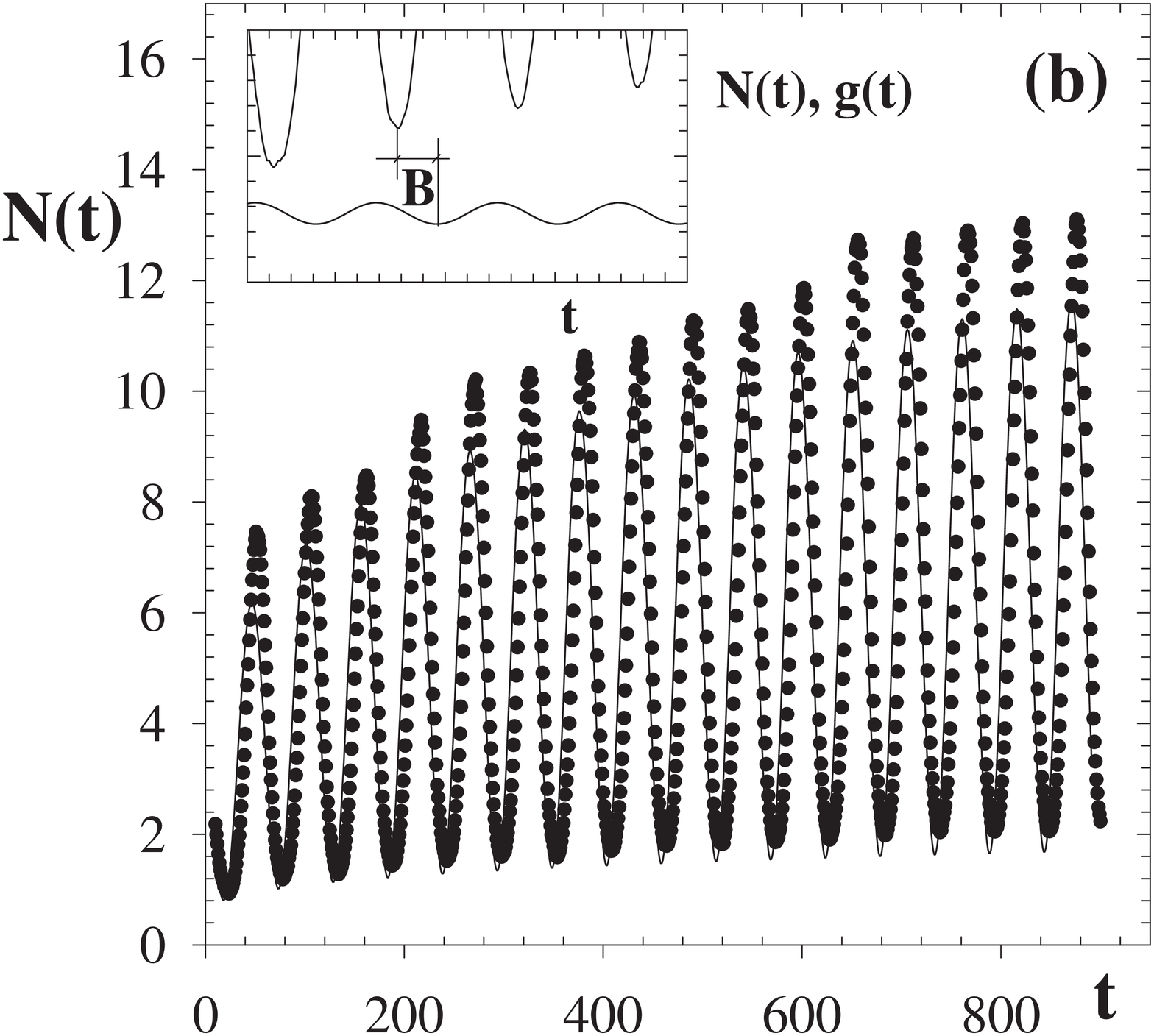,width=17cm}} Figure 2 (b) \newpage
\newpage
\hbox{\psfig{figure=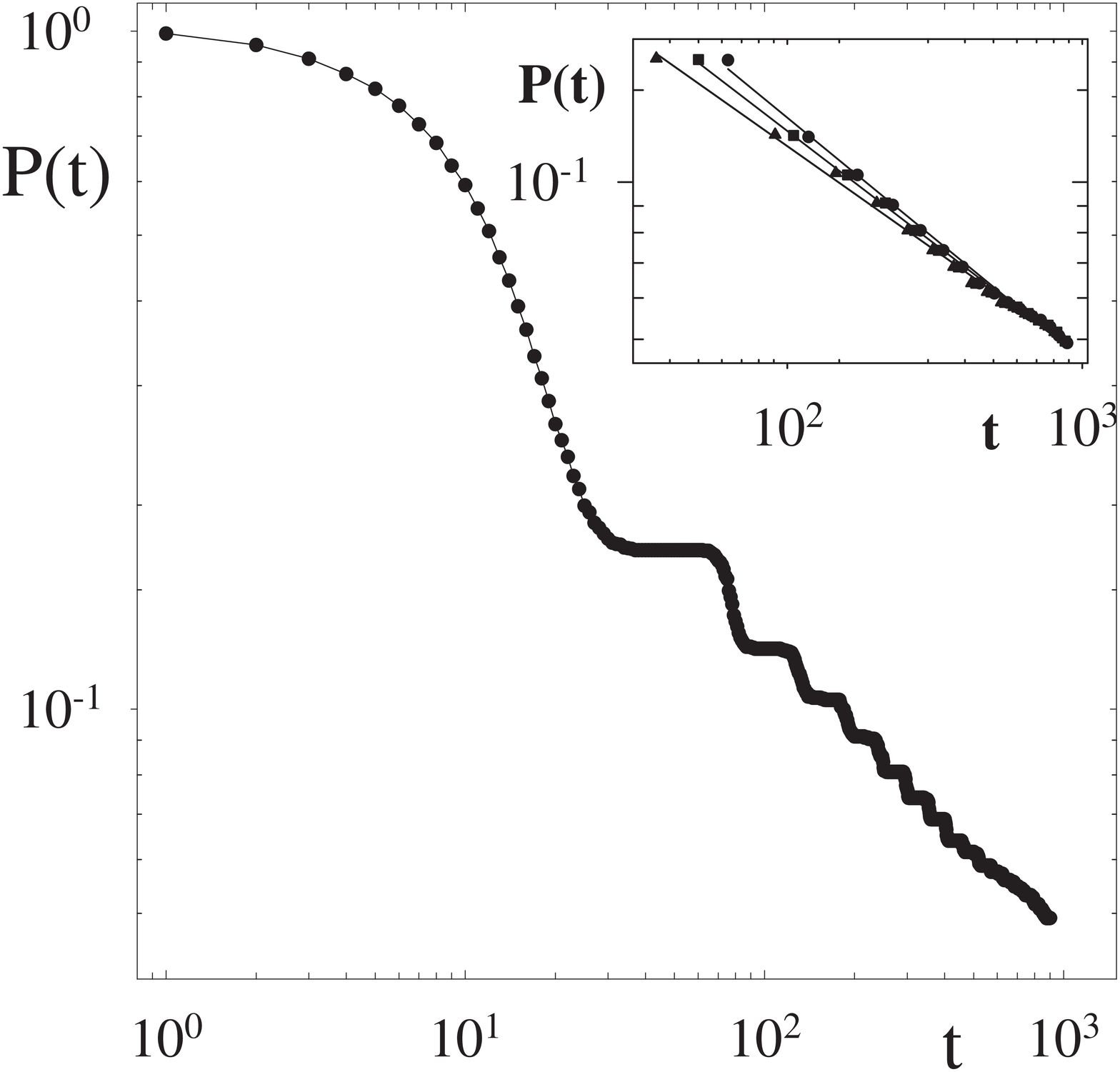,width=17cm}} Figure 3 \newpage
\hbox{\psfig{figure=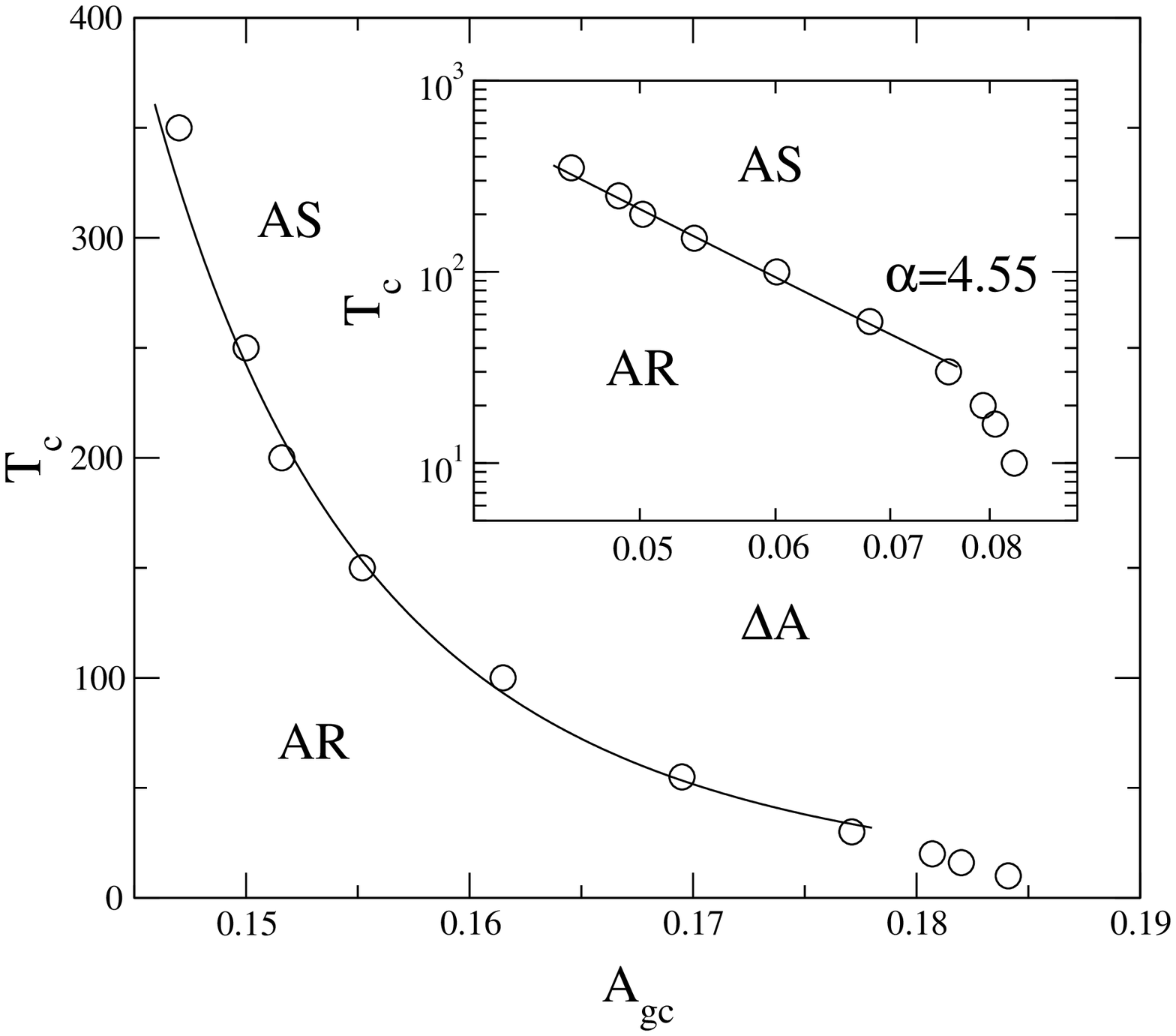,width=17cm}} Figure 4 \newpage
\hbox{\psfig{figure=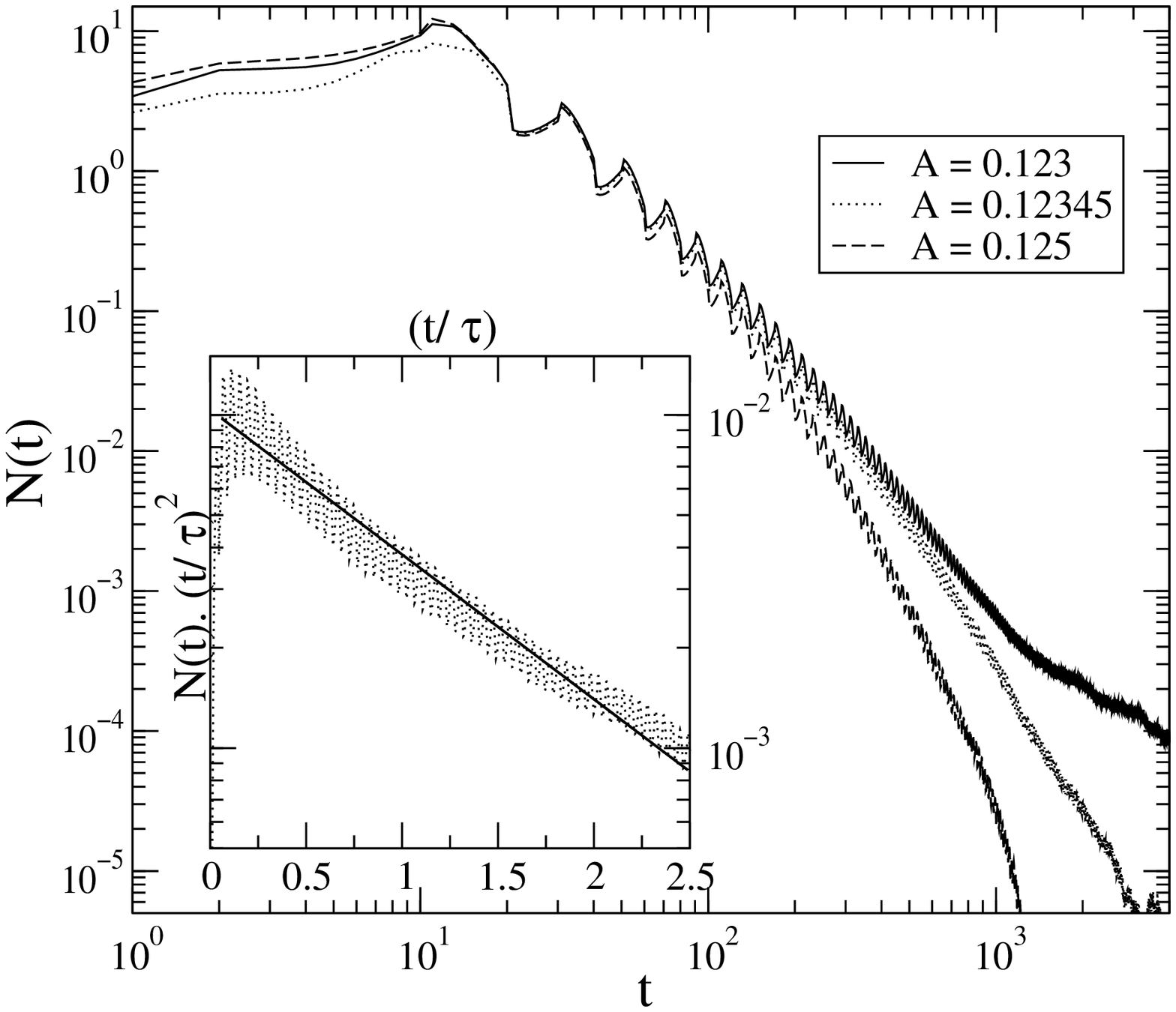,width=17cm}} Figure 5 \newpage
\hbox{\psfig{figure=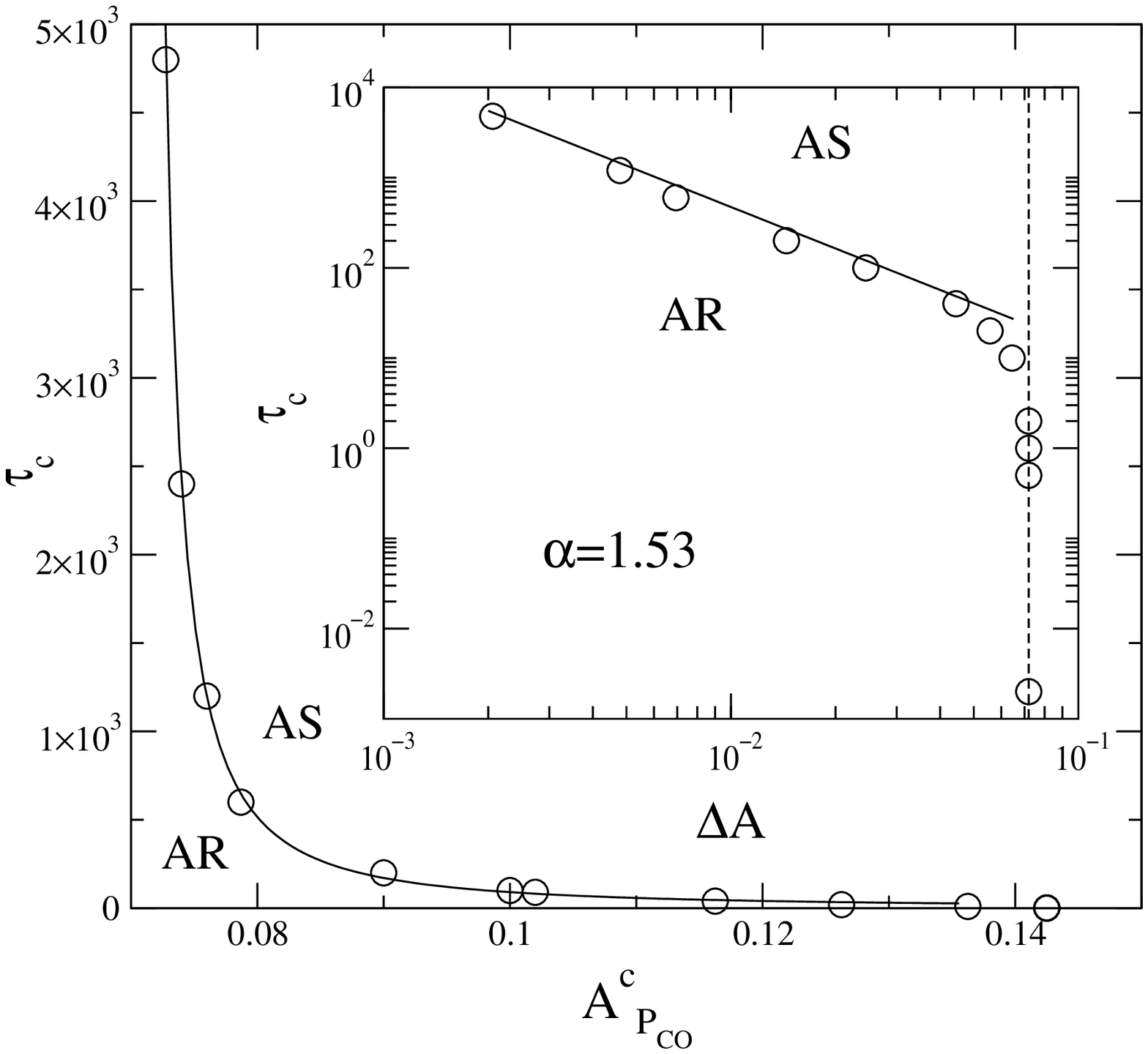,width=17cm}} Figure 6 \par}

\end{document}